\documentstyle[aps,prl,multicol,epsf,amssymb]{revtex}

\begin{document}

\bibliographystyle{prsty}
\draft
\preprint{ }
\title{Electronic Density of States of Atomically Resolved Single-Walled
Carbon Nanotubes: Van Hove Singularities and End States}

\author{Philip Kim, Teri W. Odom, Jin-Lin Huang, and Charles M. Lieber
}

\address{Harvard University, 12 Oxford Street, Cambridge, MA 02138}

\maketitle

\begin{abstract}
The electronic density of states of atomically resolved single-walled
carbon nanotubes have been investigated using scanning tunneling
microscopy.  Peaks in the density of states due to the one-dimensional
nanotube band structure have been characterized and compared with the
results of tight-binding calculations.  In addition, tunneling
spectroscopy measurements recorded along the axis of an
atomically-resolved nanotube exhibit new, low-energy peaks in the
density of states near the tube end. Calculations suggest that
these features arise from the specific arrangement of carbon atoms that
close the nanotube end.

\end{abstract}

\begin{multicols}{2}
\narrowtext


The electronic properties of single-walled carbon nanotubes (SWNTs) 
are currently the focus of considerable interest \cite{R1}.  According 
to theory \cite{R2,R3,R4}, SWNTs can exhibit either metallic or 
semiconducting behavior depending on diameter and helicity.  Recent 
scanning tunneling microscopy (STM) studies of SWNT \cite{R6,R5} have 
confirmed this predicted behavior, and have reported peaks in the 
density of states (DOS), Van Hove singularities (VHS), that are 
believed to reflect the 1D band structure of the SWNTs.  A detailed 
experimental comparison with theory has not been carried out, although 
such a comparison is critical for advancing our understaning of these 
fascinating materials.  For example, chiral SWNTs have unit cells that 
can be significantly larger than the cells of achiral SWNTs of similar 
diameter, and thus chiral tubes may exhibit a larger number of VHS 
than achiral ones\cite{R7}.  Recent theoretical work \cite{R8,R9} 
suggested, however, that semiconducting (or metallic) SWNTs of similar 
diameters will have a similar number of VHS near the Fermi level, 
independent of chiral angle.  In addition, the electronic properties 
of localized SWNT structures, including end caps, junctions and bends 
\cite{R10,R11,R12,R13}, which are essential to proposed device 
applications, have not been characterized experimentally in atomically 
resolved structures.

In this Letter, we report STM investigations of the electronic 
structure of atomically resolved SWNTs and compare these results with 
tight-binding calculations.  Significantly, we find that the VHS in 
the DOS calculated using a straight-forward zone-folding approach 
agree with the major features observed in our experiments.  We have 
observed new peaks in the local DOS (LDOS) at an end of a metallic 
SWNT and compared these results to calculations.  This analysis 
suggests that the new peaks can be associated with a specific topology 
required to cap the SWNT. The implications of these results and 
important unresolved issues are discussed.

Experimental procedures and are described elsewhere in detail 
\cite{R5,R14}.  In brief, SWNT samples were prepared by laser 
vaporization \cite{R15}, purified and then deposited onto a Au 
(111)/mica substrate.  Immediately after deposition, the sample was 
loaded into a UHV STM that was stabilized at 77 K; all of the 
experimental data reported in this Letter was recorded at 77 K. 
Imaging and spectroscopy were measured using etched tungsten tips with 
the bias ($V$) applied to the tip.  Spectroscopy measurements were 
made by recording and averaging 5 to 10 tunneling current ($I$) versus 
$V$ ($I$-$V$) curves at specific locations on atomically resolved 
SWNTs.  The feedback loop was open during the $I$-$V$ measurement 
while the setpoint was the same as that of imaging.  The tunneling 
conductance, $dI/dV$, was obtained by numerical differentiation.

\setcounter{figure}{0}
\begin{figure}[b]
\epsfxsize=60mm \centerline{\epsffile{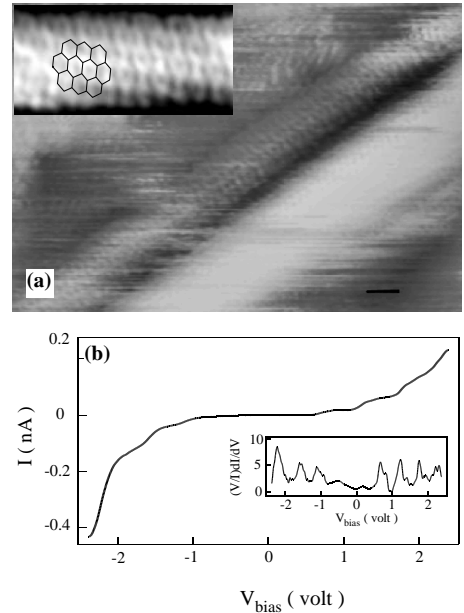}} \vskip 4mm
\caption{ (a) STM image of SWNTs recorded at $I=0.12$ nA and
$V=550$ mV. Tunneling spectra were taken on the isolated upper
tube.  The black scale bar is 1 nm.  The inset shows an atomic
resolution image of this tube. A portion of a hexagonal lattice is
overlaid to guide the eye. (b) $I$-$V$ data recorded on the SWNT
in (a). The inset shows the normalized
conductance,$(dI/dV/(I/V))$.} \label{fig1}
\end{figure} \vskip -5mm

An atomically resolved STM image of several SWNTs is shown in Fig.\ 
\ref{fig1}(a).  The upper isolated SWNT rests on the Au surface and is 
on the edge of a small rope that contains around 10 nanotubes.  Below 
we concentrate our analysis on this individual SWNT. The diameter and 
chiral angle measured for this tube were 1.35 $\pm $ 0.1 nm and -20 
$\pm$ $1 ^{\circ}$, respectively.  These values are consistent with 
$(13,7)$ and $(14, 7)$ indices \cite{R5,R14}, where $(13,7)$ and 
$(14, 7)$ are expected to be metallic and semiconducting respectively.  
The $I$-$V$ data exhibits metallic behavior with relatively sharp, 
stepwise increases at larger $|V|$ (Fig.\ \ref{fig1}(b)).  The $I$-$V$ 
curves have a finite slope, and thus the normalized conductance 
$(V/I)(dI/dV)$, which is proportional to the LDOS, has appreciable 
non-zero value around $V=0$ as expected for a metal (Fig.\ 
\ref{fig1}(b) inset).  This suggests that the $(13, 7)$ indices are 
the best description of the tube (we address this point further 
below).  At larger $|V|$, several sharp peaks are clearly seen in 
$dI/dV$ and $(V/I)(dI/dV)$ vs, $V$.  These peaks were observed in four 
independent data sets recorded at different positions along this 
atomically-resolved tube (but not on the Au(111) substrate), and thus 
we believe these reproducible features are intrinsic to the SWNT. We 
attribute these peaks to the VHS resulting from the extremal points in 
the 1D energy bands \cite{R16}.

\begin{figure}
\epsfxsize=60mm \centerline{\epsffile{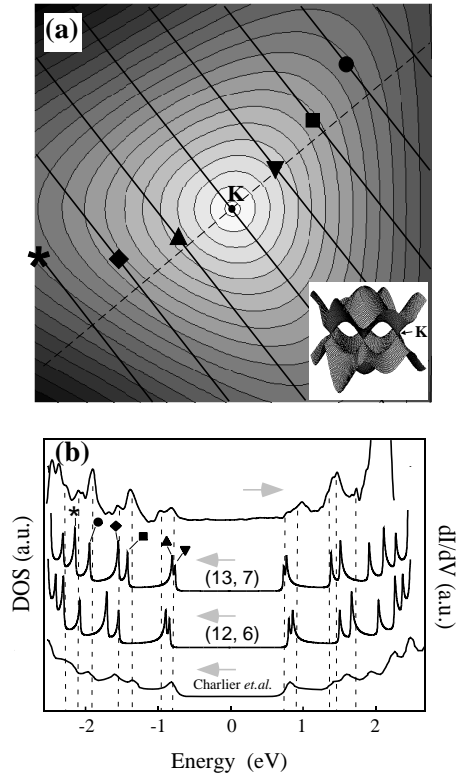}} \vskip 3mm 
\caption{(a) Energy dispersion of the $\pi$-band of graphene sheet 
near the $\bf {K}$.  The solid lines correspond to $(13,7)$ 1D bands 
obtained by the zone-folding.  Symbols are located at the positions 
where VHS occur in these 1D bands.  The inset depicts a three 
dimensional view of graphene $\pi/\pi^{*}$ bands.  (b) Comparison of 
DOS obtained from our experiment (upper curve) and $\pi$-only tight 
binding calculation for $(13,7)$ SWNT (second curve from top).  The 
broken vertical lines indicate the positions of VHS in the tunneling 
spectra after consideration of thermal broadening convolution.  The 
symbols correspond to the VHS shown in (a).  The calculated DOS of 
$(12,6)$ and an independent calculation for a $(13, 7)$ tube [9] are 
included for comparison.}
\label{fig2}\end{figure}

The availability of spectroscopic data for atomically-resolved 
nanotubes represents a unique opportunity for comparison with theory.  
In this regard, we have calculated the band structure of a $(13, 7)$ 
SWNT using the tight-binding method.  If only $\pi$ and $\pi^{*}$ 
orbitals are considered, the SWNT band structure can be constructed by 
zone-folding the 2D graphene band structure into the 1D Brillouin zone 
specified by the $(n, m)$ indices \cite{R1}.  Fig.\ \ref{fig2}(a) 
shows the graphene $\pi$ band structure around the corner point ($\bf 
{K}$) of the hexagonal Brillouin zone.  For the metallic $(13,7)$ 
tube, the degenerate 1D bands which cross $\bf {K}$ result in a finite 
DOS at the Fermi level.  Note that the energy dispersion is isotropic 
(circular contours) near $\bf {K}$, and becomes anisotropic (rounded 
triangular contours) away from $\bf {K}$.  Therefore, the first two 
VHS in the 1D bands closest to $\bf {K}$ (depicted by $\blacktriangle 
$ ,$\blacktriangledown $) have a smaller splitting in the energy due 
to the small anisotropy around K, while the next two VHS (depicted by 
$\blacksquare$, $\blacklozenge $) have a larger splitting due to 
increasing anisotropy.  If the energy dispersion were completely 
isotropic, both sets of peaks would be degenerate.  Values for the 
hopping integral, $V_{pp\pi}$, reported in the literature range from 
ca.  2.4 to 2.9 eV \cite{R1,R2,R5,R6,R8,R21}.  We used a value 
of 2.5 eV determined from previous measurements of the energy gap vs. 
diameter \cite{R5}.

Our STS data shows relatively good agreement with the DOS for a $(13, 
7)$ tube calculated using the zone-folding approach (Fig.\ 
\ref{fig2}(b)).  The agreement between the VHS positions determined 
from our $dI/dV$ data and calculations are especially good below the 
Fermi energy ($E_{F}$) where the first seven peaks correspond well.  
Deviations between experiment data and calculations are larger above 
than below $E_{F}$.  The observed differences may be due to the band 
repulsion, which arises from the curvature-induced hybridization, or 
surface-tube interaction that were not accounted for in our 
calculations.  Detailed {\it ab initio } calculation \cite{R17} have 
shown that the effect of curvature induced by hybridization is much 
higher in $\pi^{*}/\sigma^{*}$ than $\pi/\sigma$ orbitals.  Bands 
above $E_{F}$ are thus more susceptible to the hybridization effect, 
and this could explain the greater deviations between experiments and 
calculation that we observe for the empty states.  In the future, we 
believe that comparison between experiment and more detailed 
calculations should help (a) to resolve such subtle but important 
points, and (b) to understand how inter-tube and tube-substrate 
interactions affect SWNT band structure.

 \vskip -3mm

\begin{figure}
\epsfxsize=65mm \centerline{\epsffile{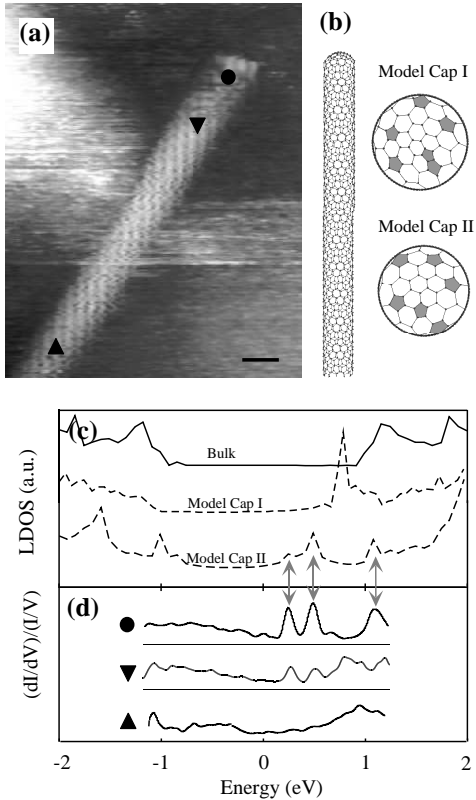}} \vskip 5mm
\caption{ (a) STM image of the end of a SWNT with $I=0.15$ nA and
$V= 750$ mV. The scale bar is 1 nm, and symbols ($\bullet$,
$\blacktriangledown$\&$\blacktriangle$) correspond to the
locations where the tunneling spectra in (d) were
recorded. (b) A model $(13, -2)$ SWNT with two different cap
structures; the pentagons in the caps are shaded gray.  (c) LDOS
obtained from a tight-binding calculation on capped $(13,-2)$
tubes. The solid and dashed curves correspond to the calculated
bulk DOS and end DOS of cap I and cap II, respectively. (d)
Experimental tunneling spectra from the end ($\bullet$), near the
end ($\blacktriangledown$), and far from end ($\blacktriangle$).
Similar features in $\bullet$,$\blacktriangledown$ and cap II are
highlighted by gray arrows.} \label{fig3}
\end{figure} \vskip -1mm

In addition, we have compared these results to a recent $\sigma+\pi$ 
calculation for a $(13,7)$ SWNT \cite{R9}and a $\pi$ only calculation 
for a closely related set of indices.  The bottom curve shown in Fig.\ 
\ref{fig2}(b) is adopted from \cite{R9}, and was obtained using 2s 
and 2p orbitals.  Although the direct comparison is difficult due to 
an large broadening, all peaks within $\pm$ 2 eV match well with our 
$\pi$-only calculation.  This comparison suggests that curvature 
induced hybridization is only a small perturbation within the 
experimental energy scale ($|V| < 2$ V) for the $(13, 7)$ tube.  We 
have also investigated the sensitivity of the DOS to $(n, m)$ indices 
by calculating the DOS of the next closest metallic SWNT to our 
experimental diameter and angle;that is, a $(12, 6)$ tube.  
Significantly, the calculated VHS for this $(12, 6)$ tube deviate much 
more from the experimental DOS peaks than in the case of the $(13, 7)$ 
tube(Fig.\ \ref{fig2}(b)).  We believe that this analysis not only 
substantiates our assignment of the indices in Fig.\ \ref{fig1}(a), 
but more importantly, demonstrates the sensitivity of detailed DOS to 
subtle variations in diameter and chirality.

Lastly, we have also investigated the electronic structure of the ends 
of atomically-resolved SWNTs.  Analogous to the resonant or localized 
surface states in bulk crystals \cite{R18}, resonant or localized 
states are expected at the ends of nanotubes \cite{R10}.  In 
accordance with Euler's rule, a capped end should contain six 
five-membered carbon rings (pentagons).  The presence of these 
topological defects can cause dramatic changes in the LDOS near the 
end of a nanotube \cite{R19}.  Previous STM studies of multi-walled 
nanotubes \cite{R13} reported localized states at the ends of these 
tubes, although the atomic structure of the tubes was not resolved.  
To the best of our knowledge, STM studies of the ends of SWNT have not 
been reported.  Fig.\ \ref{fig3}(a) shows an atomically-resolved image 
of the end of an isolated SWNT that has also been characterized 
spectroscopically.  The rounded structure exhibited in this and 
bias-dependent images (e.g., insets Fig.  4(a)) suggests strongly that 
the end is closed, although the atomic structure cannot be obtained 
since the tube axis is parallel to the image plane.  These images 
enable us to assign the nanotube $(13, -2)$ indices (the left-handed 
counterpart to an $(11, 2)$ tube).  The expected metallic behavior of 
the $(13, -2)$ tube was confirmed in $(V/I)(dI/dV)$ data recorded away 
from the end ($\blacktriangle$ in Fig.\ \ref{fig3}(a),(d)).  
Significantly, spectroscopic data recorded at and close to the SWNT 
end ($\bullet$\&$\blacktriangledown$ in Fig.\ \ref{fig3}(d)) show two 
distinct peaks at 250 mV and 500 mV that decay and eventually 
disappear in bulk DOS recorded far from the tube end 
($\blacktriangle$).  The peaks were observed in 10 independent data 
sets recorded at the tube end, and are very reproducible.

To investigate the origin of these new features, we carried out 
tight-binding calculations for a $(13, -2)$ model tube with different 
end caps (Fig.  \ \ref{fig3}(b)).  All the models exhibit a bulk DOS 
far from the end (solid curve in Fig.\ \ref{fig3}(c));however, near 
the end, the LDOS show pronounced differences from the bulk DOS:two or 
more peaks appear above the $E_{F}$, and these peaks decay upon 
moving away from the end to the bulk.  (Fig.\ \ref{fig3}(b)) shows two 
representative cap models.  These models were chosen to illustrate the 
relatively large peak differences for caps closed with isolated vs.  
adjacent pentagons.  These topological configurations are not unique, 
although additional calculations show that other isolated (adjacent) 
pentagon configuration have similar LDOS. Significantly, the LDOS 
obtained from the calculation for cap II shows excellent agreement 
with the measured LDOS at the tube end, while cap I does not (Fig.\ 
\ref{fig3}(c-d)).  The positions of the two end LDOS peaks as well as 
the first band edge of cap II match well with those from the 
experimental spectra.These results suggest that the topological 
arrangement of pentagons is responsible for the observed localized 
features in the experimental DOS at the SWNT end, and are thus similar 
to conclusions drawn from measurements on multi-walled nanotubes that 
were not atomically-resolved \cite{R13}.

\begin{figure}
\epsfxsize=90mm \centerline{\epsffile{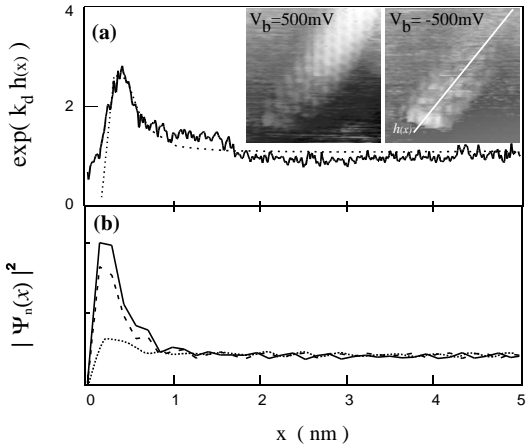}} \vskip 6mm 
\caption{(a) (inset) STM images recorded at different voltages on the 
SWNT end in Fig.\ \ref{fig3}.  The white indicates the $h(x)$ cross 
section.  The solid line in (a) corresponds to $\exp(k_{d} h(x))$, 
where $k_{d} = 2$ \AA$^{-1}$, and the dotted line is the integrated 
LDOS (0-500 meV) from our calculation.  (b) The solid ,broken, and 
dotted lines depict wavefunction probability (arb.  units), 
$|\psi_{n}(x)|^{2}$, of cap II as a function of position $x$ for 
eigenenergies of 500, 250, and 320 meV, respectively.}
\label{fig4}
\end{figure} \vskip 0mm

The nature of the DOS peaks at the nanotube end was further 
investigated using bias dependent STM imaging.  At the bias of strong 
DOS peak, -500 mV,the tip-nanotube separation $h(x)$ decays with 
increasing $x$, where $x$ is the distance from the tube end (inset, 
Fig.\ \ref{fig4}(a)).  As indicated in Fig.\ \ref{fig4}(a), 
$\exp(k_{d}h(x))$, which is proportional to the integrated LDOS 
\cite{R20}, sharply increases around the capped end of the tube and 
then decays with a characteristic length scale ca.  1.2 nm.  Our 
tight-binding calculation suggests that this decay can be attributed 
to resonant end states.  Wavefunctions whose eigenenergies correspond 
to the LDOS peaks (250, 500 meV) decay exponentially from the end into 
the bulk but retain a finite magnitude (Fig.\ \ref{fig4}(b));this type 
of decay is a signature of a resonant state \cite{R18}.  Note that 
$h(x)$ does not decay at $V$ far from the resonance (e.g.  Fig.\ 
\ref{fig3}(a)) nor do wavefunctions whose eigenenergies are away from 
the end LDOS peaks decay with $x$ (Fig.\ \ref{fig4}(b)).  Resonant end 
states in metallic tubes could serve an important function in 
electronic devices by improving the contact between nanotubes and 
electrodes.

In summary, we have reported sharp VHS in the DOS of 
atomically-resolved SWNTs using STM, and have compared these data to 
tight-binding calculations for specific tube indices.  A remarkably 
good agreement was obtained between experiment and $\pi$-only 
calculations, although deviations suggest that further work will be 
needed to understand fully the band structure of SWNTs in contact with 
surfaces.  Pronounced peaks in the LDOS were also observed at the end 
of an atomically-resolved metallic SWNT. Comparison of these data with 
calculation suggests that the topological arrangement of pentagons is 
responsible for the localized features in the experiment.  Such end 
states could be used to couple nanotube effectively to electrodes in 
future nanotube-based devices.

We thank R.E. Smalley (Rice) and C.-L. Cheung (Harvard) for SWNT
samples. T.W.O. acknowledges predoctoral fellowship support from the
NSF, and C.M.L. acknowledges support of this work by the National
Science Foundation (DMR-9306684).

\end{multicols}


\begin{references}
\bibitem{R16}Broad features observed in $(V/I)/dI/dV$ near $V 
=$ 0 do not correspond to significant changes in the DOS; the 
variation in $dI/dV$ is small compared to that of the VHS occurring at 
the band edges.  These small variations in the DOS could be attributed 
to the interaction between tube and Au substrate or adjacent tubes.
\bibitem{R20}If the wave function normal to 
the tube axis decays exponentially with a constant inverse decay 
length $k_d$, $\int_{0}^{eV}dE\ LDOS(x,E) \sim \exp(k_d h(x))$.  $k_d$ 
was obtained from experimental current versus distance measurements.
\end{references}
\end{document}